\documentclass[aps,twocolumn,amsmath,amssymb,showpacs,prb]{revtex4}
\usepackage{epsf}

\newcommand{\etal}{{\it et al.}}

\begin{document}

\title{Pseudogaps in Nested Antiferromagnets}
\author{C. P\'{e}pin$^1$ and M. R. Norman$^2$}
\affiliation{
$^1$SPhT, L'Orme des Merisiers, CEA-Saclay, 91191 Gif-sur-Yvette,
France\\
$^2$Materials Science Division, Argonne National Laboratory, Argonne,
Illinois 60439}
\date{\today}
\begin{abstract}
We analyze the fluctuation corrections to magnetic ordering in the case of
a 3D antiferromagnet with flat Fermi surfaces, as physically realized in
the case of chromium, and find that they are insufficient to produce a
quantum critical point.  This implies that the critical point observed in
vandium doped chromium is due to a loss of nesting.  We
also derive the fermion self-energy in the paramagnetic phase and find that
a pseudogap exists, though its magnitude is significantly reduced as compared
to the spectral gap in the ordered state in the limit where the latter is small in
comparison to the Fermi energy.
\end{abstract}
\pacs{75.30.Fv, 71.10.Hf, 78.20.Bh}

\maketitle

The subject of magnetic quantum critical points has sparked much interest
in the physics community.  In the case of heavy fermion metals, this interest
has largely been motivated by the observation that Fermi liquid theory breaks
down in the vicinity of such critical points \cite{GREG}.  At present, the
theory behind this breakdown is not well understood because of the strong
coupling nature of the Kondo lattice problem \cite{JPCM}.

Recently, Yeh \etal \cite{YEH} have studied the more straightforward case
of vanadium doped chromium.  Chromium is the classic example of a spin density
wave magnet driven by Fermi surface nesting \cite{RMP}. Upon doping with
vanadium, the N\'{e}el temperature, $T_N$, is rapidly suppressed to zero.  Motivated
by the speculation that the Hall number may jump at a magnetic quantum
critical point due to Fermi surface rearrangement \cite{JPCM}, the authors
of Ref.~\onlinecite{YEH} studied the Hall conductivity and indeed identified
such a jump.  Moreover, they found a strong temperature dependence of the
Hall number, which they speculated was due to the presence of a pseudogap
near the critical point.  The corresponding signature of this pseudogap
has been looked for as a spin gap in the dynamic susceptibility, but so
far results are inconclusive \cite{HAYDEN}.

The simplicity of the case of vanadium doped chromium obviously begs for a theoretical
treatment.  Recently, it has been shown that the jump of the Hall number
can be understood as due to the sudden removal of flat parts of the Fermi
surface upon magnetic ordering \cite{HALL}.  The presence of flat Fermi
surfaces obviously points to the possibility of a pseudogap, given the
quasi-1D nature of the fermion dispersion \cite{LEE}.

In this paper, we consider the flat Fermi surface model of chromium originally
proposed by Shibatani \etal \cite{FLAT} where the Fermi surface is
approximated as a cube.  We find, in agreement with
earlier work \cite{HASE}, that fluctuation corrections are less singular than
in a non nested antiferromagnet, and thus are insufficent to drive the N\'{e}el
temperature to zero.  Rather, the critical temperature must be driven to zero
by loss of Fermi surface nesting \cite{RICE}.  This is consistent with recent
pressure data \cite{NEW}, which indicate scaling exponents near the critical point
for the N\'{e}el temperature and Hall number in agreement with analytic
results based on a curved Fermi surface \cite{HALL}.
Moreover, we evaluate the
fermion self-energy in the paramagnetic phase, and indeed find a pseudogap in
the spectral function.  This pseudogap, though, scales as
$T_N^3/E_F^2$ for $T_N \ll E_F$, where
$E_{F}$ is the Fermi energy, and thus is likely to be too small to be responsible
for the strong temperature variation of the Hall
number \cite{YEH}.  Instead, the observed variation may be due to the
temperature dependence of the inelastic scattering rate, as observed in
other transition metals \cite{PA2}.  On the other hand, for more strongly correlated systems
where $T_N \sim E_F$, then the pseudogap does scale as $T_N$.

The polarization bubble for the flat case is identical to that of the BCS theory for
superconductivity \cite{MAKI}.  That is,
\begin{equation}
\chi_0 = N [\ln(1.14D/T) - \xi^2 q^2 + i \alpha \omega]
\end{equation}
where $D$ is the ultraviolet cut-off (bandwidth) and $N$ the density of states.
The expansion coefficient $\xi$ is isotropic
for the commensurate case ($Q=2\pi/a$ for chromium), and weakly anisotropic for the incommensurate
case \cite{MAKI} ($q$ is defined relative to the ordering wavevector, $Q$).
In linear response theory, the interacting susceptibility is
\begin{equation}
\chi_{MF} = \chi_0/(1-g\chi_0)
\end{equation}
where $g$ is the exchange interaction.  The zero of this denominator defines the
mean field transition temperature, $T_{MF} = 1.14De^{-1/gN}$.
The mean field inverse susceptibility is then
\begin{equation}
\chi_{MF}^{-1} = \frac{Ng}{\chi_0} [ \ln \frac{T}{T_{MF}} + \xi^2 q^2 - i \alpha \omega]
\end{equation}
Fluctuation corrections in the Hartree approximation give a true
inverse susceptibility of \cite{LT}
\begin{equation}
\chi^{-1} = \chi_{MF}^{-1} + b <M^{2}> \; ,
\end{equation}
 with $b$
diverging as $1/T^{2}$, $\alpha$ as $1/T$, and $\xi$ as $1/T$. The
system undergoes a transition towards antiferromagnetic order at a
temperature $T_N$ for which $\chi^{-1} (\omega=0, q=0) =0$.
 The important difference of the nesting case to a normal 3D
antiferromagnet is the temperature dependence of the expansion
coefficients.  Because of this, spin fluctuations become
relatively less important as the temperature is lowered, and thus
fluctuation effects are less singular \cite{HASE}.

To see this, we approximate $<M^{2}>$, the fluctuating staggered
moment, by its classical value ($ \omega \ll T$)
\begin{equation}
<M^{2}> = \frac{T}{2} \int \frac{d^{3}q}{(2\pi)^{3}} \chi(q,0) =
\frac{Tq_c\chi_0}{4\pi^{2}Ng\xi^{2}}
\end{equation}
The latter equality assumes that $T=T_N$.
 If $q_{c}$, the
classical cut-off, is assumed to satisfy the condition
$\Gamma(q_{c})=T$, then one can show that the classical value is
approximately equal to the true quantum mechanical value for
$<M^2>$ \cite{LT}. Here, $\Gamma$ is the frequency half width of
the dynamic susceptibility,
 which, at $T=T_{N}$ is $q_{c}=\sqrt{\alpha T}/\xi$.
 Using this, $<M^{2}>$ reduces to
\begin{equation}
<M^{2}> = \frac{T^{4}\sqrt{\alpha^{\prime}}\chi_0}{4\pi^{2}Ng\xi^{\prime 3}}
\end{equation}
where $\alpha^{\prime}=\alpha T$ and $\xi^{\prime}=\xi T$ are
temperature independent constants.  Recognizing that
$b^{\prime}=bT^{2}$ is also a temperature independent constant,
$\chi^{-1}(0,0)$ reduces to
\begin{equation}
\chi^{-1}(0,0) = \ln\frac{T}{T_{MF}} +
\frac{b^{\prime}\sqrt{\alpha^{\prime}}T^{2}\chi_0^2}
{4\pi^{2}N^2g^2\xi^{\prime 3}}
\end{equation}
That is, the correction to $T_{MF}$ goes as $T^2\ln^2 T$,
which is less singular than the $T^{3/2}$ correction for an ordinary
3D antiferromagnet \cite{MORIYA}.  Our result agrees with earlier results
of Hasegawa \cite{HASE}, though our derivation is more straightforward.

Since $T_{N}$ can never be driven to zero for a perfectly flat
Fermi surface (due to the logarithmic divergence of $\chi_{0}$),
then one might think that the quantum critical point is probably
not controlled by fluctuations \cite{FOOT1}. Rather, loss of
nesting is the likely cause of the quantum critical point. This is
consistent with band theory results \cite{HALL}, which find an
increasing mismatch of the electron and hole octahedral surfaces
as the hole doping is increased.  Such warping corrections will
cause the fluctuation corrections to cross over to the standard 3D
antiferromagnetic result near the critical point.

Having addressed the question of fluctuations, we now turn to the
question of the pseudogap.  Postulating a pseudogap in the case of
nesting is quite natural given the quasi-1D nature of the fermion
dispersion.  On the other hand, as we have seen above, despite this
quasi-1D behavior, the spin fluctuation spectrum is still 3D-like, and this
raises questions about how strong the pseudogap effect will be.  To
address this, we derive the fermionic self-energy to lowest order.
We note that this is given by a convolution of an effective interaction
$V$ with the fermion Greens function, where \cite{BERK}
\begin{equation}
V = g + g^2 \chi  + g^2 \chi/(1+g \chi_0) \simeq (3g/2) \chi/\chi_0
\end{equation}
with the first term the bare interaction, the second one from summing
a ladder series, and the third from summing a bubble series.  The most
singular part of the fermionic self-energy comes from the classical
fluctuations, and can be approximated as \cite{VILK}
\begin{equation}
\Sigma = \frac{3T}{2N} \int \frac{d^{3}q}{(2\pi)^{3}}
\frac{1}{A + \xi^{2}q^{2}}\frac{1}{\omega-\epsilon_{k+Q+q}}
\end{equation}
where $\epsilon$ is the fermionic dispersion and $A =
\ln\frac{T}{T_{MF}}+\frac{b\chi_0}{Ng}<M^2>$.  In the flat case,
$\epsilon_{k+Q+q} = -\epsilon_{k} - v q_{\parallel}$, where $v$ is
the Fermi velocity and $q_{\parallel}$ is normal to the flat
surface.  $\Sigma$ now becomes
\begin{equation}
\Sigma = \frac{3T}{2N} \int \frac{dq_{\parallel}d^{2}q_{\perp}}{(2\pi)^{3}}
\frac{1}{A + \xi^{2}(q_{\parallel}^{2}+q_{\perp}^{2})}
\frac{1}{\omega+\epsilon_{k}+vq_{\parallel}}
\end{equation}
The integral over $q_{\perp}$ gives
($\xi^{2}q_{c}^{2} = \alpha T$)
\begin{equation}
\Sigma = \frac{3T}{8\pi N\xi^{2}} \int \frac{dq_{\parallel}}{2\pi}
\ln \left ( \frac{A+\xi^{2}q_{\parallel}^{2}+\alpha T}
{A+\xi^{2}q_{\parallel}^{2}} \right )
\frac{1}{\omega+\epsilon_{k}+vq_{\parallel}}
\end{equation}
The $q_{\parallel}$ integral is convergent, and the cut-off can be taken to
infinity.  The result is
\begin{equation}
Re\Sigma = \frac{3T}{8\pi N v \xi^{2}} \left
(\tan^{-1}\frac{\xi(\omega+\epsilon_{k})}{v\sqrt{A}}
-\tan^{-1}\frac{\xi(\omega+\epsilon_{k})}{v\sqrt{A+\alpha T}}
\right )
\end{equation}
\begin{equation}
Im\Sigma = \frac{3T}{16\pi N v \xi^{2}}
\ln\frac{A+(\xi/v)^{2}(\omega+\epsilon_{k})^{2}+\alpha T}
{A+(\xi/v)^{2}(\omega+\epsilon_{k})^{2}}
\end{equation}

To understand these expressions further, we assume that $T=T_{N}$
($A$=0) where the pseudogap effect should be most pronounced.  This
yields
\begin{equation}
Re\Sigma = \frac{\bar{\Delta}^2}{2.7T_N} \left
(\frac{\pi}{2}sgn(\omega+\epsilon_{k})
-\tan^{-1}\frac{\omega+\epsilon_{k}}{2.7T_{N}} \right )
\end{equation}
\begin{equation}
Im\Sigma = \frac{\bar{\Delta}^2}{5.4T_N}
\ln(1+(\frac{2.7T_{N}}{\omega+\epsilon_{k}})^{2})
\end{equation}
where a typical energy scale ${\bar \Delta}$ is defined as
\begin{equation}
\bar{\Delta}^2=8.1T_N^4/(8\pi N v \xi^{\prime 2})
\end{equation}
These were obtained by noting that for the flat case \cite{HASE},
$\alpha^{\prime}=\pi/8$ and $\xi^{\prime}=v\sqrt{7\xi(3)/(16\pi^{2})}$,
where $\xi(3)=1.202$.  We note that $2.7T_N$ is the natural frequency
scale of the problem, and that $\bar{\Delta}$ has units of energy.

The fermion spectral function is given as
\begin{equation}
{\cal A} (k, \omega) =
\frac{1}{\pi}\frac{Im\Sigma}{(\omega-\epsilon_k-Re\Sigma)^2+(Im\Sigma)^2}
\end{equation}
In Fig.~1, we show the spectral function at the Fermi surface for two cases,
$\bar{\Delta}= 2.7T_N$ and $0.27T_N$.  In the first case
(strong coupling limit), the spectral gap is approximately equal to
$\bar{\Delta}$.  In the second case (more appropriate for chromium as will
be seen below),
the spectral gap is significantly smaller than $\bar{\Delta}$.
In Fig.~2, we plot the spectral gap obtained from half the spectral peak
to peak
separation in Fig.~1 versus $\bar{\Delta}$.  For large $\bar{\Delta}$
(comparable to $T_N$), the spectral gap scales with
$\bar{\Delta}$, whereas for small $\bar{\Delta}$, the spectral gap
scales quadratically with $\bar{\Delta}$.

To understand these results analytically, we note that the pole of the
fermion Greens function on the Fermi surface is given by the
condition $\omega-Re\Sigma(\omega)=0$.  Let us first assume this pole
energy is of order $2.7T_N$.  Then under these conditions,
$Re\Sigma$ in Eq.~12 can be approximated as (high frequency expansion)
\begin{equation}
Re\Sigma_{high} =  \frac{\bar{\Delta}^{2}}{\omega}
\end{equation}
This expression is identical to the BCS expression for the self-energy,
and the pole can easily be seen to occur at an energy $\bar{\Delta}$.
That is, there is a spectral gap equal to $\bar{\Delta}$, and this explains
the behavior for large $\bar{\Delta}$ in Fig.~2.
By noting that the density of states for the flat case, $N$, is $1/(2\pi v a^2)$,
then
\begin{equation}
\bar{\Delta} = 6.2 T_{N}^{2}a/v
\end{equation}
Since $v/a \sim E_{F}$ ($a$ is the lattice constant), this would imply that
$T_N$ must be of order the Fermi energy, $E_F$, for this high
frequency approximation
to be valid.  This is not satisfied for chromium, since
$T_N \ll E_F$ in that case.

In the other limit, one expands the self-energy for small $\omega$, obtaining
(low frequency expansion)
\begin{equation}
Re\Sigma_{low} = \frac{\bar{\Delta}^2}{2.7T_N}(\frac{\pi}{2}sgn(\omega)-
\frac{\omega}{2.7T_N})
\end{equation}
The pole energy is then given approximately by
$\frac{\pi}{2}\bar{\Delta}^2/(2.7T_N)$.  Therefore, for the case where
$T_N \ll E_F$, then the spectral gap scales as $T_N^3/E_F^2$.

\begin{figure}
\centerline{\epsfxsize=3.4in{\epsfbox{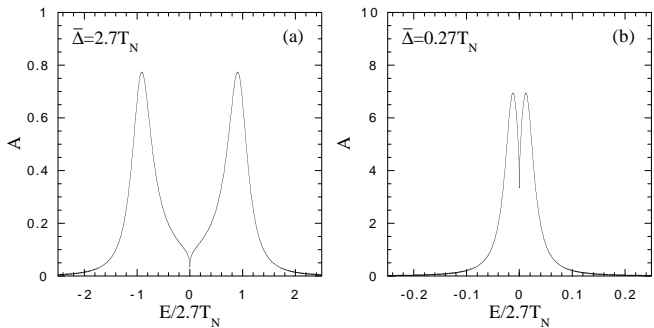}}} \caption{
Spectral function, ${\cal A}$, on the Fermi surface derived using
Eqs.~14-17 for two values of $\bar{\Delta}$.  Note differing
energy scales in the two plots. } \label{fig1}
\end{figure}

\begin{figure}
\centerline{\epsfxsize=2.0in{\epsfbox{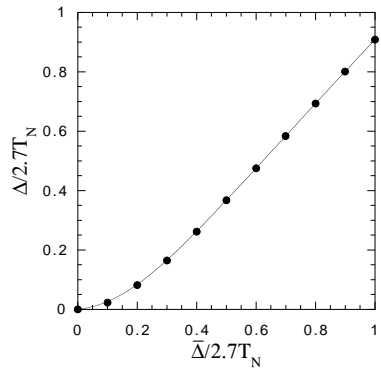}}}
\caption{
Spectral gap, $\Delta$, (half the peak to peak separation in Fig.~1) versus
$\bar{\Delta}$ derived using Eqs.~14-17.  Note quadratic behavior of
the spectral gap for small $\bar{\Delta}$ and linear behavior for
large $\bar{\Delta}$.
}
\label{fig2}
\end{figure}

\begin{figure}
\centerline{\epsfxsize=2.0in{\epsfbox{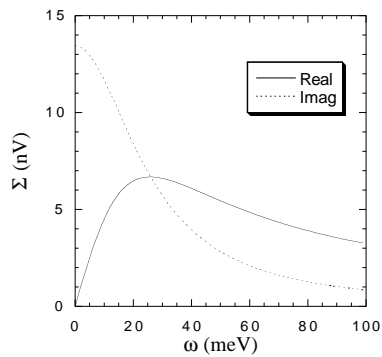}}}
\caption{
Self-energy on the Fermi surface calculated from Eqs.~12-13 using neutron scattering
parameters from Cr-V 5\%.  Although the $\omega$ scale is reasonable,
the magnitude of $\Sigma$ is so small (nanovolts) that a pseudogap does not
develop.
}
\label{fig3}
\end{figure}

Pure chromium exhibits a classic mean field transition as far as
specific heat measurements are concerned \cite{RMP}.  This implies
that it is in the weak coupling limit, consistent with the
small ratio of $T_N$ to $E_{F}$.
We can quantify this by using parameters extracted from neutron
scattering data.  In Ref.~\onlinecite{HAYDEN}, the authors use a
form for the susceptibility identical to the one employed here
\begin{equation}
\chi(q,\omega)=\frac{\chi_Q}{1+q^2/\kappa_0^2-i\omega/\omega_{sf}}
\end{equation}
By comparing to our expressions, we see that
$\kappa_0^2=A/\xi^2$ and $\omega_{sf}=A/\alpha$.
Using this, the prefactor outside the parenthesis in Eq.~12 becomes
$6T^2\kappa_0^2a^2/(\pi\omega_{sf})$ and the quantities
dividing $(\omega+\epsilon_k)$ in the $\tan^{-1}$ functions in Eq.~12 are
$2.7\sqrt{\omega_{sf}T}$ and $2.7\sqrt{\omega_{sf}T+T^2}$
respectively.  For Cr-V 5\%, $\kappa_0=0.11 \AA^{-1}$ and
$\omega_{sf}=88 meV$ for T=12K ($a = 2.88 \AA$).
We plot the resulting self-energy from Eqs.~12 and 13 in Fig.~3.
Though the $\omega$ structure of $\Sigma$ is reasonable
(looking like a damped version of Eq.~18 with a maximum in Re$\Sigma$ at 26 meV),
the value of $\Sigma$ itself (nanovolts) is far too small to cause
a pseudogap.

Based on this, we expect
only weak pseudogap effects, even in the magnetically ordered part of the
phase diagram.  We note that our derivation of the self-energy assumes
that the spin structure factor is quasi-static, a property of the renormalized
classical regime \cite{VILK}.  That is, we would not necessarily expect pseudogap
effects in the quantum critical regime.  This conclusion is bolstered
by our evalulation of quantum corrections to the self-energy, which we
do not find to be singular.

\begin{figure}
\centerline{\epsfxsize=2.0in{\epsfbox{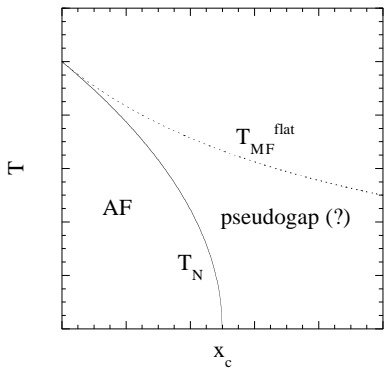}}}
\caption{
Illustration of a possible scenario for the pseudogap phase in the V doped Cr
system.  $T_{MF}^{flat}$ is the mean field temperature for a flat Fermi surface,
$T_N$ the actual transition temperature which is suppressed by loss of nesting
(warping of the Fermi surface).  The pseudogap, if it exists, should be confined to
the region between these two temperatures.
}
\label{fig4}
\end{figure}

On the other hand, in the flat model, $T_{MF}$ never vanishes as a function
of doping.  It is a loss of
nesting which leads to the quantum critical point.  Therefore, it is possible that a pseudogap
exists for all dopings which satisfy $T_N < T < T_{MF}$ .  This is illustrated
in Fig.~4.
Still, we expect
that although pseudogap effects are possible near the quantum critical point of
vanadium doped chromium, they are likely to be weak.
They could perhaps be best searched for by photoemission, which sees the
spectral gap quite easily in pure chromium \cite{ARPES}.

This begs the question of what is responsbile for the strong
temperature dependence of the Hall number observed by Yeh \etal \cite{YEH} which
occurs even for dopings far beyond the quantum critical point.
The Hall number is temperature
dependent in transition metals such as Cu, which can be attributed to the temperature
dependence of the electron-phonon scattering rate \cite{PA2}.  For the vanadium doped
chromium case, this would be consistent with the $T^3$ dependence of the
resistivity \cite{YEH}, which points to the prevalence of electron-phonon
effects.
Moreover, the experimental Hall number \cite{YEH} is in excess of the
paramagnetic band theory value \cite{HALL} for temperatures above 150 K,
again indicating the presence of an inelastic scattering contribution (which could
be of magnetic origin as well).
Calculation of the $T$ dependence of the
Hall number, though, is technically challenging \cite{PA2} since it
involves going beyond the Boltzmann approximation, so we do not consider
this further here.

On the other hand, our results do indicate a large pseudogap in
the strong coupling limit.  We note that there are examples of quantum critical
points where Fermi liquid theory is known to break down, and where
nesting may be playing an important role, such as in the case of the
bilayer ruthenate $Sr_{3}Ru_{2}O_{7}$ \cite{SRO}.  Moreover,
magnetic incommensurability is seen in most quantum critical
heavy fermion systems, such as Au doped $CeCu_{6}$ \cite{SCHR}.  It
is possible that the results presented here, which were derived for
the case where a strong Fermi surface rearrangment takes place at the
quantum critical point, are quite relevant for these systems.  Based on this,
we suggest that pseudogap effects be searched for in heavy fermion
quantum critical systems.

We would like to thank the authors of Ref.~\onlinecite{YEH}, particularly
Tom Rosenbaum, for communicating their work to us prior to publication.
This work was supported by the U. S. Dept. of Energy, Office of Science,
under Contract No. W-31-109-ENG-38.  CP would like to thank the
hospitality of ANL, and MN that of the SPhT, while this work was in
progress.


\begin{thebibliography}{99}
\bibitem{FOOT1}
In the presence of disorder, $T_{N}$ can be driven to zero in the flat
case since nesting is smeared out.  This is unlikely to be the case for
chromium, in that $T_{N}$ can be driven to zero by pressure instead of
doping \cite{RMP,NEW}.
\end{thebibliography}
\end{document}